\documentclass[reprint,aps,amsmath,amssymb,nofootinbib,twoside,prd,showkeys,superscriptaddress]{revtex4-1}
\pdfoutput=1
\usepackage{verbatim}
\usepackage{comment}
\usepackage{graphicx}
\usepackage[T1]{fontenc}
\usepackage{epsfig}
\usepackage{bm}
\usepackage{tensor}
\usepackage{amssymb}
\usepackage{float}
\usepackage{amsmath}
\usepackage{subfigure}
\usepackage{dcolumn}
\usepackage[utf8]{inputenc}
\usepackage{cancel}
\usepackage[colorlinks]{hyperref}
\usepackage[usenames,dvipsnames]{color}
\hypersetup{
     breaklinks=true,
    pdfstartview={FitH},    
    colorlinks=true,       
    linkcolor=blue,          
    citecolor=red,        
    filecolor=magenta,      
    urlcolor=blue,           
    anchorcolor=green,      
    linktocpage=true
}

\def\doi{http://doi.org}

\newcommand{\HCd}{\mathcal{H}}

\def\HCdt0{\tilde{\HCd}_{0}}

\usepackage{multirow}
\usepackage{graphicx}
\usepackage{tikz}

\newlength\imagewidth
\newlength\imagescale

\newcommand{\affcam}{DAMTP, Centre for Mathematical Sciences, University of Cambridge, Wilberforce Road, Cambridge CB3 0WA, United Kingdom}
\newcommand{\affcamast}{Kavli Institute of Cosmology (KICC), University of Cambridge, Madingley Road, Cambridge, CB3 0HA, United Kingdom}

\newcommand{\affFran}{Institut de Physique Theorique, Universite Paris-Saclay, CEA, CNRS, F-91191 Gif-sur-Yvette Cedex, France}
\newcommand{\cern}{CERN, Theoretical Physics Department, Geneva, Switzerland.}

\begin{document}

\title{Multi-scale Constraints on Scalar-Field couplings to Matter: 
\\ The Geodetic and Frame-Dragging Effects}
\author{David Benisty}
\email{db888@cam.ac.uk}
\affiliation{\affcam}\affiliation{\affcamast}
\author{Philippe Brax}
\email{philippe.brax@ipht.fr}
\affiliation{\affFran}\affiliation{\cern}
\author{Anne-Christine Davis}
\email{ad107@cam.ac.uk}
\affiliation{\affcam}\affiliation{\affcamast}
\begin{abstract}
The impact of light scalars coupled conformally and disformally to matter on the geodetic and frame-dragging (FD) precessions is calculated. For larger frequencies the disformal interaction becomes increasingly relevant. We use several satellite experiments and Pulsar time of arrival (ToA) measurements to derive bounds on the couplings, combining the Gravity Probe B, LARES, LAGEOS and GRACE results with pulsar timings. Forecasts for future constraints on the conformal and the disformal couplings based on the GINGER experiment, i.e. a future measurement of the Sagnac effect on Earth, the motion of $S$-stars around the galactic centre and future pulsar timing observations are presented.    
\end{abstract}
\maketitle
\section{Introduction}
Einstein's theory of General Relativity (GR) has been probed using gravitational tests in  the solar system and within galactic environments  - most recently by the Gravitational Wave detection. Despite its resounding success  in describing the present day Universe, GR's connection to both early and late time phenomena is problematic. Examples can be found in the  early Universe with the  unexplained nature of the Big Bang Singularity or the  initial conditions that would  give rise to the standard model of Big Bang cosmology. Another area where  GR does not give a definite answer is the nature of the accelerated expansion of the Universe~\cite{Perlmutter:1998np,Weinberg:1988cp}. The famous and  mysterious component of the Universe called  dark energy does not have a consistent microscopic model within Quantum Field Theory and GR~\cite{Peebles:2002gy}. Both these reasons have motivated the study of extensions of GR in different astrophysical systems~\cite{Baker:2019gxo,2022npjMG849B}.

The class of theories  we are broadly interested in here  can obey current observational constraints and mimic GR when they exhibit screening effects whereby the modifications to GR  are hidden at small scales. Observed large scales are mostly unaffected by the modification of gravity apart from some interesting and small effects on the growth of structure. More specifically, in this paper we focus on theories where the conformal coupling to matter depends on the environment, i.e. the distribution of matter around the considered objects, e.g. satellites in the Earth's atmosphere or pulsars in the Milky Way\footnote{For description of these models and a comparison with scalarisation, see \cite{Benisty:2022lox}.}. The most general coupling of a scalar field to matter is obtained via the conformal and disformal terms as appearing in the Jordan frame metric as shown by Bekenstein in ~\cite{Bekenstein:1992pj}. Such theories give rise to fifth forces generically, which are subject to strict limits by solar system tests of GR \cite{Bertotti:2003rm}. Consequently,  the effect of fifth forces need to be screened in the solar system, giving rise to screened modified gravity models. Such models can be screened via various mechanisms \cite{Khoury:2003aq, Brax:2004qh,Hinterbichler:2010es,Vainshtein:1972sx,Damour:1994zq,Brax:2010gi}. All rely in different ways  on the environment and are such that the fifth force is screened in the solar system, i.e. the theory evades all solar system test, and  can give rise to modifications to GR on cosmological scales. 
Similarly the disformal coupling to matter gives rise to modifications to GR which can be constrained from  the solar system to collider physics ~\cite{Brax:2018bow,Brax:2015hma}. This results in constraints on the disformal coupling to matter \cite{Koivisto:2008ak,Zumalacarregui:2010wj,Koivisto:2012za,vandeBruck:2013yxa,Brax:2013nsa,Neveu:2014vua,Sakstein:2014isa,Sakstein:2014aca,Desmond:2019ygn} and forecasts for future satellite experiments \cite{Vagnozzi:2021quy,Brax:2020hkc}.

We analyse the effect of both the conformal and disformal couplings on the geodetic and FD precessions.  Current astrophysical observations are used to constrain the coupling strengths in the conformal and disformal cases. In practice, we consider scalar-tensor theories of gravity and their effect on the precession  frequency of rotating gravitating objects \cite{Brax:2020vgg}. The Jordan metric $g^J_{\mu\nu}$ is related to the Einstein metric thanks to  the metric transformation \cite{Bekenstein:1992pj,Sakstein:2014isa,vandeBruck:2015ida,Koivisto:2012za}:
\begin{equation}
    {g}_{\mu\nu}^{(J)} =  \left(1+{2\beta}\frac{\phi}{m_{\rm Pl}}\right) g_{\mu\nu}^{(E)} + \frac{2}{m_{\rm Pl}^2 \Lambda^2} \partial_{\mu}\phi\partial_{\nu}\phi
\end{equation}
where ${g}_{\mu\nu}^{\left(J\right)}$ is the metric in Jordan frame and $g_{\mu\nu}^{\left(E\right)} $ is the metric in the Einstein frame. $\beta$ is conformal coupling strength with matter, which will depend on the environment, and  $\Lambda$ is the disformal coupling strength with matter.We have denoted by $m_{\rm Pl}$ the Planck mass. Notice that $\beta$ is dimensionless and $\Lambda$ has dimension of mass.  Recently light scalar fields have also been suggested as possible candidates for dark matter \cite{Hui:2016ltb}. The coupling of such dark matter fields to matter is also crucial for their dynamics and their eventual detection ~\cite{Brax:2017xho,Trojanowski:2020xza,Brax:2020gqg}.

These theories can be tested using gravitational methods as shown by earlier studies which focused on two bodies in an orbital motion  ~\cite{Maheshwari:1980gmr,Damour:1991rd,Buonanno:1998gg,Damour:1988mr,Damour:1999cr,Benisty:2022txp,Benisty:2022idt,Benisty:2022ive} 
a well studied example in GR, from which similar properties can be inferred for light scalars with conformal and disformal couplings ~\cite{Brax:2012ie,Brax:2013uh,Zhang:2017srh,Brax:2018bow,Davis:2019ltc,Brax:2019tcy,Liu:2017xef,Shibata:2022gec,Benisty:2022lox}.

\begin{figure}
    \centering
\includegraphics[width=0.44\textwidth]{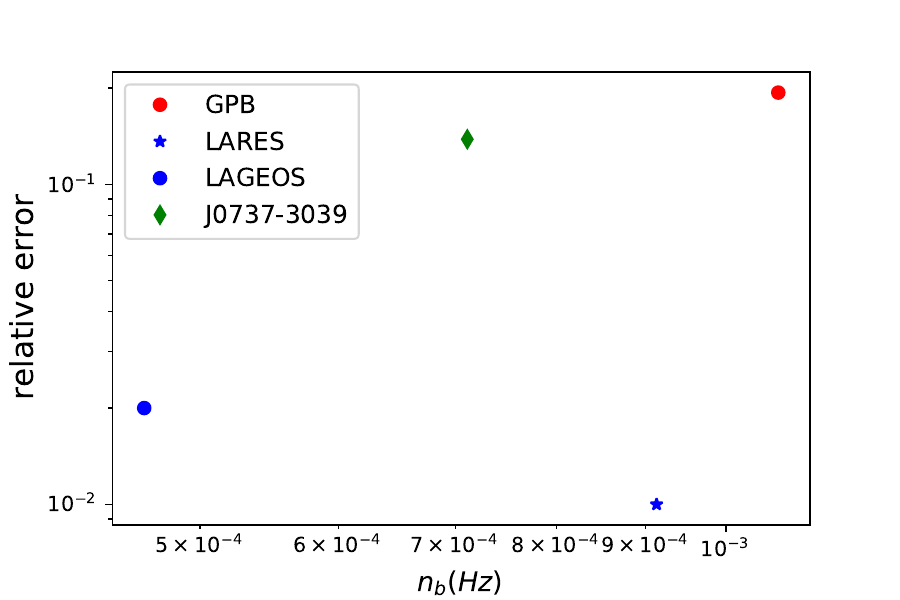}
\caption{\it{ Comparison of the geodetic and the frame dragging measurements from different data sets that are discussed in this paper: we give the accuracy of the experiment vs. the orbital frequency.}}
    \label{fig:overview}
\end{figure}

We will focus on tests of the geodetic and the FD effects. The geodetic effect (or de-Sitter) follows from  the curvature of spacetime predicted by general relativity, and the way it acts on a vector carried along with an orbiting body \cite{10.1093/mnras/77.2.155,Bagchi:2017sjo}. The FD, or Lense-Thirring, effect is one of the main predictions of Einstein’s theory of gravitation in the limit of weak field and slow motion, i.e. it represents a tiny relativistic precession of the orbital plane of a satellite produced by the angular momentum of the primary object 
\cite{Schiff:1960gh,Barker:1975ae,1975ApJ...199L..25B,1979GReGr..11..149B,Mashhoon:1984fj,2007GReGr..39.1735P,Yunes:2013dva}. 
The difference between the geodetic and FD effects is that the de-Sitter one is due simply to the presence of a central mass, whereas FD precession is due to the rotation of the central mass. The total precession is calculated by combining the de-Sitter precession with the FD precession. The precessions read in GR
\begin{subequations}
 \begin{equation}
\Omega_{dS} = \frac{3 G n_b }{2a c^2\left(1-e^2 \right)} \frac{m_2 \left( 4 m_1 + 3 m_2\right)}{\left(m_1 + m_2 \right)^{4/3}},    
\end{equation}   
 \begin{equation}
\Omega_{FD} = \frac{3 G S}{2a c^2\left(1-e^2 \right)^{3/2}}, 
\end{equation}
\end{subequations}
where $G$ is the Newtonian Constant, $n_b = 2 \pi/P_b$ is the orbital frequency, $a$ is the semi-major axis, $c$ is the speed of light, $e$ is the eccentricity, $S$ is the spin of the central body and $m_{1,2}$ are the masses of the bodies. The directions of the vectors are
\begin{equation}
\vec{\Omega}_{dS} = \Omega_{dS}\vec{k}, \quad \vec{\Omega}_{FD} = \Omega_{FD}(\vec {s}' - 3 (\vec{k} \cdot \vec{s}') \vec {k}) ,
\end{equation}
where where $\vec{k} = \vec{J}/J$ is the unit vector along the orbital angular momentum, $\vec{J}$ is the orbital angular momentum and $\vec{s}$ is the spin vector of the companion body.

Ref~\cite{Brax:2018bow} has extended the leading order calculations in GR to include both conformal and disformal couplings to matter in scalar-tensor theories applied to the two body problem in the Post Newtonian expansion. This enables one to  test these theories  in new regimes, such as the galactic centre where stars and the supermassive black hole orbit around each other ~\cite{Benisty:2021cmq}. \cite{Benisty:2022lox} derives the corresponding  Post Keplerian Parameters and the influence of the conformal and disformal couplings.  The higher derivative nature of the disformal interaction is parameterised using the dimensionless quantity $\epsilon_\Lambda$ which relates the disformal coupling interaction to the frequency
\begin{equation}
\epsilon_\Lambda = \left(\beta \cdot n_b / \Lambda\right)^2/\left(1-e^2\right)^3,
\label{eq:epsilonLambda}
\end{equation}
where $n_b = 2\pi/P_b$ is the frequency of the motion and $P_b$ is the period of the motion. $\epsilon_\Lambda$ parameterises the contributions of the disformal interaction to the Post Keplerian Parameters (PKP). As we will see, this parameter also appears naturally in the geodetic and in the FD terms.  Fig.~\ref{fig:overview} compares the experiment that we discuss in this paper and in particular the relative error of the precession rate vs. the orbital period. For larger $n_b$ with lower errors the bound on the disformal coupling is expected to be the strongest.

Notice that the bounds derived here from satellite experiments are not as strong as the ones obtained in particle physics \cite{Brax:2015hma} or even with pulsar timings \cite{Benisty:2022lox}. The particle physics and pulsar timing results involve energies and environments which differ from the ones in the solar system tests. As such the results presented here complement the known bounds on the conformal and disformal couplings and are environment specific\footnote{This is particularly relevant for models where the couplings are environment dependent such as symmetrons \cite{Hinterbichler:2010es} for instance.}.

The structure of this paper is as follows: Section \ref{sec:conDisInter} calculates the geodesic and the frame-dragging effects for scalar tensor theories with conformal and disformal interactions. Section~\ref{sec:satellites} describes the constraints on the interactions from current and future satellites experiments. Section~\ref{sub:GINGER} discusses the GINGER experiment in details. Section \ref{sec:pulsars} considers the binary pulsars and their current and future constraints. Section \ref{sec:center} discusses the possible detection of effects in the galactic centre, and finally section \ref{sec:results} summarises  current and future results.

\section{Light scalars interacting with matter}
\label{sec:conDisInter}

\subsection{The interactions}
The dynamics of gravity interacting with a massless scalar field are described by
\begin{equation}
S=\int d^4x \sqrt{-g}\left ( m_{\rm Pl}^2\frac{R}{2}-\frac{1}{2} g^{\mu\nu}\phi_{,\mu}\phi_{,\nu}\right) +S_m(\psi_i, g_{\mu\nu}^J),
\end{equation}
where matter fields are denoted by $\psi_i$ and  their action by $S_m$. In the following we will take the matter action to be the one of point-like particles and the scalar potential to be vanishing. In particular, we take the mass of the scalar field to be vanishing. In practice, we assume that the Compton wavelength $1/m$ of the scalar field, where $m$ is the scalar mass, is much larger than the scales we are considering. Typically, scalar effects are Yukawa-suppressed by an exponential term $e^{-mr}$ where $r$ is the distance to a gravitational source and therefore no scalar effects are expected if $r\lesssim m^{-1}$. We focus on the regime where the dynamics of the scalar are not Yukawa-suppressed and thus will simply take the mass to be vanishing. This will provide  an appropriate description of the dynamics of macroscopic objects like neutron stars as long as finite size effects can be neglected. This setting can also apply to screened models where  the scalar field between massive objects is assumed to be very light and the coupling to matter reduced by the appropriate screening mechanism in order to pass solar system tests of gravitation. In particular, we consider that $\beta$ depends on the environment, i.e. it could be different around a pulsar and in the solar system, see \cite{Benisty:2022lox} for a more thorough discussion, hence the bounds that we will deduce are specific to the given environment of each of the considered binary systems.  Here we take $\beta$ to be universal, i.e. it does not depend on the nature of the objects but only of their environment. The case of non-universal couplings proportional to the inverse compactness of the objects is highly relevant to the screened phase of modified gravity models \cite{Benisty:2022lox}. One of the main effects of taking non-universal couplings for different objects would be the dipolar power loss for binary system which would not vanish and would affect the time evolution of  binary pulsars. This would of course affect the parameter space of the models. This is left for future work.

In the following, we will be interested in precession effects for bound orbits, e.g. binary systems. The metric and the scalar field will be treated in perturbation theory where several parameters govern the corresponding expansions. First of all, we will expand in the conformal coupling $\beta^2$ which is considered to be small. We will also consider the Post-Newtonian expansion (PN) in the small velocities $v^2 \simeq GM/r$ for bound orbits where $v$ is a typical speed, $M$ a typical mass and  $r$ the size of the orbit. The disformal interactions will also be taken perturbatively  in a ladder expansion \cite{Brax:2018bow}. This expansion is valid when the ladder parameter
\begin{equation}
\epsilon_L = \left(\frac{v}{c} \right)^2 \frac{G M }{r^3\Lambda^2} 
\end{equation}
is small. When this is not the case, a summation of the ladder contributions must be performed as in \cite{Davis:2021oce}. Here we will consider situations where $v/c \ll 1$ and the Newtonian potentials are $GM/r \ll 1$ on the orbits of the binary systems. Moreover $\Lambda$ plays the role of a Ultra-Violet (UV) cut off of the theory above which higher order derivative corrections to the disformal coupling are expected. As a result we focus on the low energy regime $r\Lambda \gg 1$ implying that the ladder parameter is safely lower than unity. Finally, we only consider spin effects at leading order.

\begin{figure*}
    \centering
\includegraphics[width=0.45\textwidth]{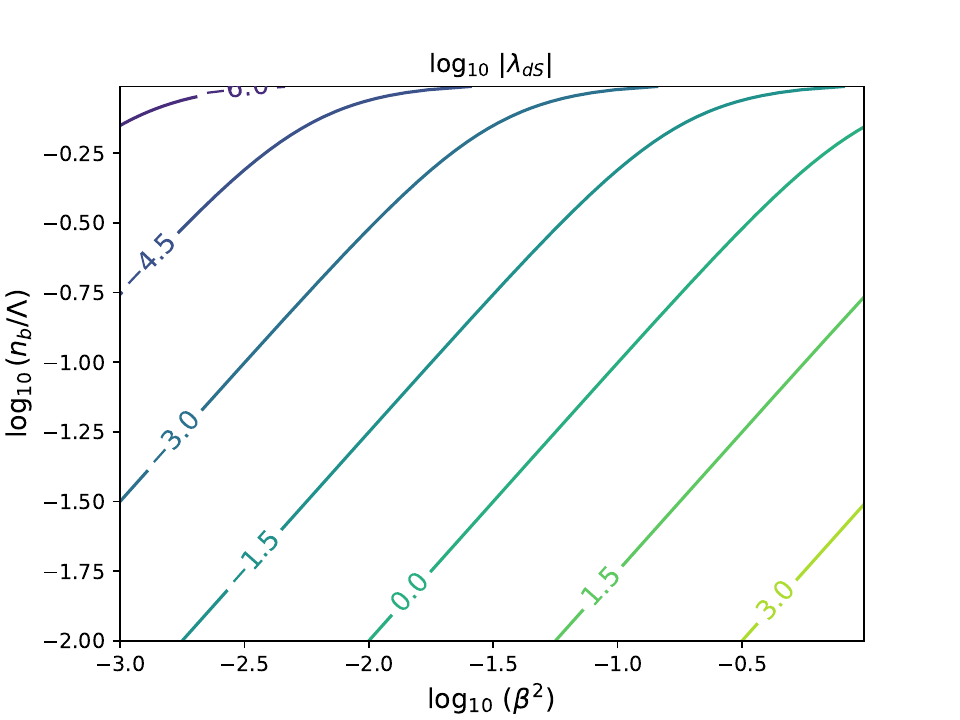}
\includegraphics[width=0.45\textwidth]{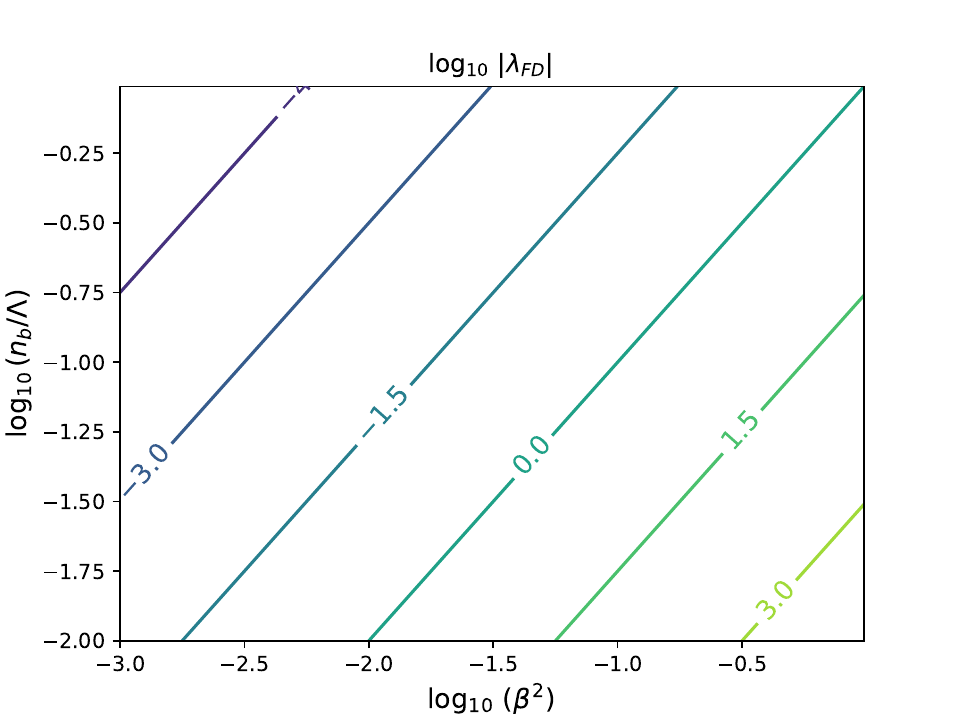}
\caption{\it{Contour plots of the geodetic (left) and the frame-dragging effects (right). The contour shows the logarithm of the modification of the geodetic and the frame-dragging effects for different values of the conformal and disformal interactions, with $e \rightarrow 0$ and $m_1 \gg m_2$. }}    \label{fig:mogratio}
\end{figure*}

\subsection{Spin Precession of coupled scalars}
General Relativity predicts the rotational drag of inertial frames in  the vicinity of a rotating object. The precession caused by this rotational drag in the motion of a rotating object is the FD effect. Gyroscopes are test objects which are  sensitive to  this effect. A gyroscope is nothing but a rotating test particle with spin and we can derive the precession of the spin vector using the Mathisson–Papapetrou–Dixon equations as done in \cite{Brax:2020vgg,Brax:2021qqo}. The evolution equation for the dynamics of a spinning test body is given by
\begin{equation}
    \dot{p}_{\mu} =-\frac{1}{2} R^J_{\mu \nu\rho\sigma}u^{\nu}S^{\rho \sigma}
\end{equation}
and
\begin{equation}
    \dot{S}^{\mu\nu}= 2 p^{[\mu}u^{\nu]}
\end{equation}
where $p^{\mu}$ is the $4$-momentum, $u^{\mu}$ is the $4$-velocity, $S^{\mu\nu}$ is the spin angular momentum, $R^{\mu\nu\rho\sigma}_J$ is the Riemann tensor of the Jordan frame metric, i.e. the metric felt by the spinning particle. The time derivative is taken along the particle world-line as $\frac{d}{d\tau}= u^\mu \nabla_\mu$ where $\tau$ is the proper time.

We can eliminate the gauge degrees of freedom by imposing the Spin Supplementary Condition (SSC)~\cite{Pryce:1948pf,Hanson:1974qy,Brax:2020vgg}, $    S^{\mu\nu}p_{\nu} \approx 0$. At leading order in the spin we can write the first step of an iteration scheme as follows~\cite{Chicone:2005jj}
\begin{equation}
    \dot{u}_{\mu} = 0+...
\end{equation}
and
\begin{equation}\label{eq:spindot}
    \dot{S}_{\mu\nu} =0 +...
\end{equation}
where the neglected  terms  are of higher order in the spin. This implies that the spin is parallel transported. This also simplifies the SSC to 
\begin{equation}
    S^{\mu\nu}u_{\nu} \approx 0
\end{equation}
which is preserved along the particle world-line $\frac{d}{d \tau}( S^{\mu\nu}u_{\nu}) =0$ guaranteeing that the correct number of degrees of freedom is preserved. Indeed, 
the spin tensor can be projected onto  the Pauli-Lubanski vector as
\begin{equation}
    S_{\mu}=-\frac{1}{2}\epsilon_{\mu\nu\rho\sigma}u^{\nu}S^{\rho\sigma}.
\end{equation}
This encodes the three independent degrees of freedom which remain after imposing the SSC. We will focus on  the components of the spin vector in a comoving tetrad frame. For the weakly rotating bodies that are considered here  we have
\begin{equation}
    g^J_{\mu\nu}=\eta_{\mu\nu}+h^J_{\mu\nu}
\end{equation}
and the tetrad vectors 
\begin{equation}
    e^0_{\hat{i}}=\left(1+\frac{1}{2}v^2 + h_{00}^J\right)v_i+\frac{1}{2}h_{ij}^Jv^j+h_{0i, J}+\mathcal{O}(v^5)
\end{equation}
\begin{equation}
    e^i_{\hat{j}}=\delta^i_j +\frac{1}{2}\left(v^iv_j-h^i_{j,J}\right)+\mathcal{O}(v^4)
\end{equation}
where $e^{\mu}_{\hat{0}}=u^{\mu}$, $v^{\mu}=u^{\mu}/u_0$ and the hatted indices apply to the local frame. This allows one to project onto the comoving frame. In this case, the spin evolution equation in the comoving frame \eqref{eq:spindot} becomes,
\begin{equation}
    \frac{d}{d\tau}S_{\hat{i}}=\left( {\Omega}\times  {S}\right)_{\hat{i}}
\end{equation}
and the  resulting  precession  \cite{Brax:2020vgg,Brax:2021qqo}
\begin{equation}
    \Omega^{\hat{i}}_A =\frac{1}{4}\epsilon^{ijk}\left(v_A^j\partial_k{h}_{00}^J-2v^{\mu}_A\partial_j{h}_{k\mu}^J\right)
\end{equation}
as a function of the velocity  of the spinning particle $A$. This can be separated, at leading order in the velocity, as 
\begin{equation}
\vec \Omega_A^G= \frac{1}{4}\vec v_A \times \vec \nabla {h}_{00}^J
\end{equation}
which would correspond to the geodetic precession in GR. Similarly, one can introduce the analogue of the FD as it would appear in GR 
\begin{equation}
\vec \Omega_A^{FD}= -\frac{1}{2} v^{0}_A (\vec \nabla \times \vec A_J)
\end{equation}
where we have introduced the vector $ A_J^i= h_{0i}^J$.
When the scalar field couples to matter, the metric is influenced by the conformal and disformal terms. Decomposing the metric and the scalar field around a flat background
\begin{equation}
    g_{\mu\nu} = \eta _{\mu\nu} +{h_{\mu\nu}}, \;\;\; \phi = 0 +\varphi,
\end{equation}
the Jordan metric  becomes
\begin{equation}
g_{\mu\nu}^J = \eta _{\mu\nu} +h^J_{\mu\nu}+2\beta \frac{\varphi}{m_{\rm Pl}}\eta_{\mu\nu} + \frac{2}{m_{\rm Pl}^2\Lambda^2} \varphi_{,\mu}\varphi_{,\nu}+...
\end{equation}
where we have introduced
\begin{equation} 
h^J_{\mu\nu}= h_{\mu\nu}+2\beta \frac{\varphi}{m_{\rm Pl}}\eta_{\mu\nu} + \frac{2}{m_{\rm Pl}^2\Lambda^2} \varphi_{,\mu}\varphi_{,\nu}+...
\end{equation}
At leading order we have for an N-body system
\begin{equation}
\varphi= -\frac{\beta}{4\pi m_{\rm Pl}} \sum_A \frac{m_A}{\vert \vec r -\vec r_A\vert}
\end{equation}
where the bodies are located at $\vec r_A$. 
For a two-body system, we find that the total precession of the first body induced by the scalar field, excluding the spin effects that we will discuss below, is proportional to the angular momentum of the body
\begin{equation}
    \ \Delta \vec\Omega^{dS}\equiv \vec \Omega^G +\vec \Omega^{FD}= \beta^2 G m^2_2 \left(-\frac{r^3 }{m_1+m_2 }+\frac{4G}{\Lambda^2r^6}\right) \vec \ell
\label{eq:LTfinal}
\end{equation}
where $r$ is the distance between the two bodies and $\vec \ell = \vec r\times \vec v$. This is a spin-orbit effect.
Notice that there are two contributions including one involving the disformal coupling. We retrieve the result of \cite{Brax:2020vgg} where
\begin{equation}
\bar c = 2 \beta, \quad \frac{\bar d}{M^4 }= \frac{2}{m_{\rm Pl}^2 \Lambda^2} 
 =  \frac{4\pi G}{\Lambda^2} .   
\end{equation}
This de-Sitter effect is complemented by the Frame-Dragging effect whose origin follows from the way the spin sources the scalar field in the Klein-Gordon equation.

The FD effect due to the spins of the bodies affects the precession vector. This contribution is mediated by the scalar field via $h_{00}^J$ and its dependence on the spin of the objects which sources  the scalar field~\cite{Brax:2020vgg} via the disformal interaction. In the case of a satellite revolving around the Earth the extra precession is given by
\begin{equation}
\Delta \vec \Omega^{FD}= -\frac{4 \beta^2 G^2 m_s}{r^6 \Lambda^2} \left[\left( \vec{S}\cdot \vec{v}\right) \vec{v} - v^2 \vec{S} \right],
\end{equation}
where $\vec v$ is the speed of the satellite of mass $m_s$ and $\vec S$ is the spin vector of the Earth. This is a spin-spin effect. Notice that this term does not have the dipolar nature of the usual FD effect. This term differs from the de Sitter effect which depends on the angular momentum of the bodies. Here the FD effect is proportional to the spin of the bodies.

Surprisingly and contrary to GR, the de-Sitter effect coming from the scalar field follows from both the curvature in the Jordan frame  $h_{00}^J$ and the gravito-magnetic field $\vec A_J$. The FD effect itself follows from the curvature $h^J_{00}$ sourced by the spin of the rotating bodies in the Klein-Gordon equation of the scalar field. However, we can still separate the de Sitter and FD contributions from the fact that the former depends on the angular momentum of the system and the latter on the  spins.

\subsection{Spin-Orbit precession}

In the following, we will compare the corrections to the spin-orbit and spin-spin precessions using diverse projections of the time-averaged precession vectors over a period. Defining by $\langle \cdot \rangle$ this averaging procedure, we obtain
\begin{equation}
\langle \Delta \Omega^i\rangle = \beta^2 G m_2^2 \, \epsilon^{ijk}\left( -\frac{1}{m_1+m_2} L^{jk}_3 + \frac{4G}{\Lambda^2} L^{jk}_6 \right)
\end{equation}
where we have defined the tensors
\begin{equation}
L^{ij}_n= \langle \frac{r^i v^j}{r^n}\rangle.
\end{equation}
The angular momentum always points in the normal direction $\vec n $  to the orbital plane and we obtain
\begin{equation}
\langle \Delta \vec \Omega\rangle= \Delta \Omega~ \vec n
\end{equation}
where
\begin{equation}
\Delta \Omega= \beta^2 G m_2^2 \left(-\frac{1}{m_1 + m_2} L_3 + \frac{4G}{\Lambda^2} L_6 \right).
\end{equation}
We set the normal vector $\vec{n}$ along the z-axis and using $L_n$ for the time-average of the magnitude of the angular momentum (i.e. $L^{ij} = \epsilon^{ijk} L_k$). We can choose the orbital plane to be at $z=0$. As a result, the component of the angular momentum along the $z$ direction is given by
\begin{equation}
L_n =\langle \frac{r^x v^y-r^yv^x}{r^n}\rangle.
\end{equation}
The Keplerian solution reads 
\begin{align*}
\begin{split}
\frac{r}{a}= \frac{1-e^2}{1+e \cos \theta}, \quad \dot \theta = n_b \frac{ \sqrt{(1-e^2)}}{\left(r/a\right)^2},
\end{split}
\end{align*}
where $\theta$ is the true anomaly. Using this, we can get the velocities as a function of  $\theta$
\begin{eqnarray}
&&
v^x= \dot r \cos \theta+ r\dot \theta \sin\theta = \frac{n_b a}{\sqrt{1-e^2}}\sin \theta \left(2 e \cos\theta+1\right),
\nonumber \\
&&v^y= \dot r \sin \theta- r\dot \theta \cos\theta,
\nonumber = - \frac{n_b a}{\sqrt{1-e^2}} \left(e \cos 2 \theta+\cos\theta\right). \\
\end{eqnarray}
We obtain the value of $L_n$ by the average over the unperturbed trajectories
\begin{align*}
\begin{split}
\left \langle \mathcal{A} \right \rangle = \frac{1}{2\pi} \int_0^{2\pi} d\theta \frac{(1-e^2)^{3/2}}{(1+e c)^2} \mathcal{A},
\end{split}
\end{align*}
that gives
\begin{equation*}
L_3 = \frac{n_b}{a}\frac{1}{1 - e^2}, \quad L_6 = \frac{n_b}{a^4} \frac{1 + 3 e^2 + \frac{3 }{8}e^4}{\left(1 - e^2\right)^4}.
\end{equation*}
In the case of circular orbits we have the explicit expression
\begin{equation}
L_n= \frac{v}{r^{n-1}}= \frac{n_b a}{r^{n-1/2}}.
\end{equation}
The general solution for the spin-orbit precession contribution gives
\begin{eqnarray}
&&
\langle \Delta \Omega_{dS} \rangle  =-  \frac{m_2^2}{\left( m_1 + m_2\right)^{1/2}} \frac{G^{3/2}}{a^{5/2}\left(1 - e^2\right)} 
\nonumber \\ 
&& \times \left[\beta^2  - \epsilon_\Lambda \left(1 + 3 e^2 + \frac{3 }{8}e^4 \right)\right],
\end{eqnarray}
where $\epsilon_\Lambda$ quantifies the disformal strength, as in Eq.~\ref{eq:epsilonLambda}. The disformal strength is affected by the frequency of the orbital motion, where higher frequencies give larger disformal contributions. This follows from the higher derivative nature of the disformal interaction. In the corresponding astronomical units
\begin{eqnarray}
&&
\Delta \Omega_{dS} = - \frac{T^{2/3}_{\odot} n_b^{5/3}}{1-e^2} \frac{m_2^2}{(m_1+m_2)^{4/3}} 
\nonumber \\ 
&&\times \left[\beta^2  - \epsilon_\Lambda \left(1 + 3 e^2 + \frac{3}{8} e^4\right)\right].
\end{eqnarray}
where $m\equiv m/M_\odot$ is the mass of the object in solar mass units  and $T_{\odot} = G M_{\odot}/c^2$ is the scale of the period. In order to compare the contribution of the conformal and the disformal coupling to the GR one,
we calculate the ratio between the de-Sitter precession with the conformal and disformal interactions $\lambda_{dS}:= \Delta\Omega_{\text{dS}}/\Omega_{\text{dS}}^{(GR)}$
to get
\begin{equation}
\lambda_{dS} = -\frac{2 m_1}{3 m_1 + 4m_2}\left[\beta^2  - \epsilon_\Lambda \left(1 + 3 e^2 + \frac{3 }{8}e^4 \right)\right].
\label{eq:modDS}
\end{equation}
In the following section, we will analyse the complementary contribution coming from the FD effect. The left panel of fig~\ref{fig:mogratio} shows a contour plot of the geodetic effect. The contour shows the logarithm of the modification of the geodetic effect for different values of conformal and disformal interactions, with $e \rightarrow 0$ and $m_1 \gg m_2$. For the limit $m_2 \ll m_1$ the pre-factor in the previous expression becomes $2/3$ and for the case $m_2 \sim m_1$ it is  $2/7$. Finally, we notice that a non-vanishing eccentricity only enhances the contribution from the disformal coupling. 

Let us comment on the PN corrections to this result compared to the disformal effect. In (\ref{eq:modDS}), the term in $\beta^2$ should be corrected at the next PN order by a term in $\beta^2 v^2$ where $v\ll 1$ is a typical velocity of the gyroscope. This term is negligible compared to the $\beta^2$ contribution but could compete with disformal effect in $\epsilon_\Lambda$. If $\beta^2 v^2 \ll \epsilon_\Lambda$, the disformal effect dominates over the conformal effect at the next PN order. On the other hand when $\epsilon_\Lambda \lesssim \beta^2 v^2$, the disformal effect is negligible compared to the leading $\beta^2$ contribution. In all cases, we can trust formulae like (\ref{eq:modDS}) as the next PN order in $\beta^2 v^2$ does not play a significant role.

\subsection{Frame-Dragging precession}
Similarly for the FD (or spin-spin) precession we introduce the tensor $T^{ij} = v^iv^j/r^6$ such that
\begin{equation}
\langle \Delta  \Omega_{FD}^i\rangle = -\frac{4 \beta^2 G^2 m_s}{\Lambda^2} \left[ S_j \langle T^{ij}\rangle   - \langle T \rangle S^i \right],
\end{equation}
where $T=T^i_i$ is the trace of the matrix $T^{ij}$ which reads
\begin{eqnarray}
&&\langle T^{ij}\rangle  = \frac{G M (e c +1)^6}{a^7 \left(1 - e^2\right)^7} \times
\nonumber \\ 
&&\left(
\begin{array}{ccc}
 (s+e s_2)^2 & -(2 c e+1) (c+c_2 e) s & 0 \\
 -(2 c e+1) (c+c_2 e) s & (c+c_2 e)^2 & 0 \\
 0 & 0 & 0 \\.
\end{array}
\right)\nonumber \\
\end{eqnarray}
where $\sin \theta = s$, $\cos \theta = c$, $\sin 2\theta = s_2$ and $\cos 2\theta = c_2$. The average gives
\begin{eqnarray}
&&\langle T^{ij} \rangle = \frac{G M}{2a^7 \left(1-e^2\right)^{11/2}} \times
\nonumber \\ 
&&\left(
\begin{array}{ccc}
\frac{5 e^6}{16}+\frac{41 e^4}{8}+\frac{13 e^2}{2}+1 & 0 & 0 \\
0 & \frac{7 e^6}{16}+\frac{61 e^4}{8}+\frac{19 e^2}{2}+1 & 0 \\
 0 & 0 & 0 \\
\end{array}
\right).\nonumber \\
\end{eqnarray}
For circular orbits $e\equiv 0$ this reduces to
\begin{equation}
\langle T^{ij}\rangle =\frac{G(m_1+m_s)}{2r^7}(\delta^{ij}-n^in^j),
\end{equation}
where $\vec n$ is perpendicular to the orbit. This
implies that
\begin{equation}
\langle \Delta \vec \Omega_{FD}\rangle = \frac{2 \beta^2 G^3 m_s (m_1+m_s)}{r^7 \Lambda^2} \left[\left( \vec{S}\cdot \vec{n}\right) \vec{n} +  \vec{S} \right].
\end{equation}
With the ansatz for the spin vector $\vec{S} = S \vec {z}$, we obtain the FD contribution from the disformal coupling
\begin{eqnarray}
&&
\langle \Delta \Omega^{FD}\rangle = \frac{2 \beta ^2 G^3 \left(m_1 + m_s\right) m_s S}{a^7 \Lambda ^2} 
\nonumber \\ 
&&
 \times\frac{1 +8 e^2 +\frac{51}{8} e^4 + \frac{3 }{8}e^6}{\left(1-e^2\right)^{11/2}} \sin \psi.
\end{eqnarray}
This can be rewritten as 
\begin{equation}
\langle \Delta \Omega_{FD}\rangle  = \epsilon_\Lambda \left(\frac{G m_s}{a c^2}\right) \frac{G S }{a^3} \frac{1 +8 e^2 +\frac{51}{8} e^4 + \frac{3 }{8}e^6}{\left(1-e^2\right)^{5/2}} \sin \psi.
\end{equation}
where $\psi$ is the angle between the two spin vectors. Therefore, the modification for the frame-dragging precession rate $\lambda_{FD}:= \Delta\Omega_{\text{FD}}/\Omega_{\text{FD}}^{(GR)}$ gives
\begin{equation}
\lambda_{FD} = \epsilon_\Lambda \Phi_s \frac{1 +8 e^2 +\frac{51}{8} e^4 + \frac{3 }{8}e^6}{1-e^2} \sin \psi
\label{eq:modFD}
\end{equation}
with the potential $\Phi_s = G m_s/\left(a c^2\right)$ of the satellite or the companion mass.

The functional dependence of the FD effect on the masses differs from the dS one. In particular we see that they enter now via two dimensionless parameters, i.e. $\epsilon_\Lambda$ and $\Phi_s$. The first one characterises the ladder expansion and is only sensitive to the total mass of the system whereas the second one is a characteristic of the satellite. In the dS case, the masses appear only as a dimensionless ratio $m_1/(3m_1+ 4m_2)$ which depends only on the mass ratio $m_1/m_2$. Note too  the increased sensitivity on the conformal and disformal interactions for large eccentricities. 

The right panel of fig~\ref{fig:mogratio} shows a contour plot of the FD effect. The contour shows the logarithm of the modification of the FD effect for different values of conformal and disformal interactions with $e \rightarrow 0$ and $m_1 \gg m_2$.

\subsection{The prior and the Likelihood}
We perform a full Markov-Chain Monte-Carlo (MCMC) analysis for different experiments. The parameter $\epsilon_\Lambda$ quantifies the contribution of the disformal interaction and depends on the orbital frequency of the body. Our prior is a flat prior with $\beta^2 \in [0,1]$ and $\Lambda^{-1} \in [0,n_b^{-1}]$ where $n_b$ is the orbital period of the system. We use an affine-invariant MCMC sampler for the minimisation of our likelihoods via the implementation of the open-source package $\text{Polychord}$~\cite{Handley:2015fda}. Based on ref.~\cite{Skilling:2006gxv},  the $\text{Polychord}$ estimates the evidence. One begins by drawing $400$ live points uniformly from the prior. After some iterations, the point with the lowest likelihood is replaced by a new live point drawn uniformly from the prior with the constraint. The convergence is reached when the new evidence $Z_{live}$ is some small fraction of the original one. The standard fraction of the $\text{Polychord}$ is $\epsilon < 1\%$.

The geodetic modification effect includes the conformal and the disformal couplings in two different parts, while the FD modification includes the conformal and the disformal contributions as a multiplicative factor. The FD experiments constrain the ratio $\beta/\Lambda$ directly. In order to find a lower bound on $\Lambda$, we combine the FD results with the strong bound from the Cassini experiment taken  as a Gaussian prior \cite{Bertotti:2003rm}. The bound reads 
\begin{equation}
\beta^2 = \left(2.1\pm  2.3\right) \cdot 10^{-5}, 
\end{equation}
where radio signals were sent  from the Earth to the Cassini satellite and the Shapiro time delay was analysed. In this case we take the conformal coupling $\beta$ to be the same in the binary system environment and in the solar system. In the geodetic case the bound on $\beta$ is independent of the Cassini bound. Similarly, in the analysis of Gravity probe B, the experiment is embedded in the solar system so the bounds on $\beta$ from this experiment can be compared directly to that of Cassini. 

Finally all our results depend on the ladder expansion of the disformal interaction. This requires that $\epsilon_L\ll 1$. We have checked that this is the case in our analyses. As $\epsilon_\Lambda \simeq v^2 n_b^2/\Lambda^2$ and as we impose a prior where $\Lambda^{-1} \in [ 0, n_b^{-1}]$ we see that $\epsilon_\Lambda \ll 1$ in all the cases that we consider.

The likelihood for different experiments reads:
\begin{equation}
-2\ln \, \mathcal{L} \left(\beta,\Lambda \right) = \sum_{i = 1}^{N_{PSR}} \left(\frac{\xi(\beta,\Lambda) - \xi_{ob}}{\delta \xi_{ob}}\right)^2
\end{equation}
where $\xi_{ob} \pm \delta\xi_{ob}$ is the observed precession vs. the theoretical prediction $\xi(\beta,\Lambda)$ from the modified model with its  dependence on the conformal and the disformal couplings.

\section{Satellite experiments}
\label{sec:satellites}
In this section we discuss the bounds from current and future satellite experiments. Since the periods of these systems are in the same range, we expect to get similar bounds on the disformal coupling.  As these experiments are all within the solar system, the bounds obtained here are on the couplings in this particular environment. In particular, when constraining the couplings using FD experimental results, we will complement the measurements with the Cassini bound to deduce solar system constraints on the disformal coupling.

\begin{figure}
 	\centering
\includegraphics[width=0.46\textwidth]{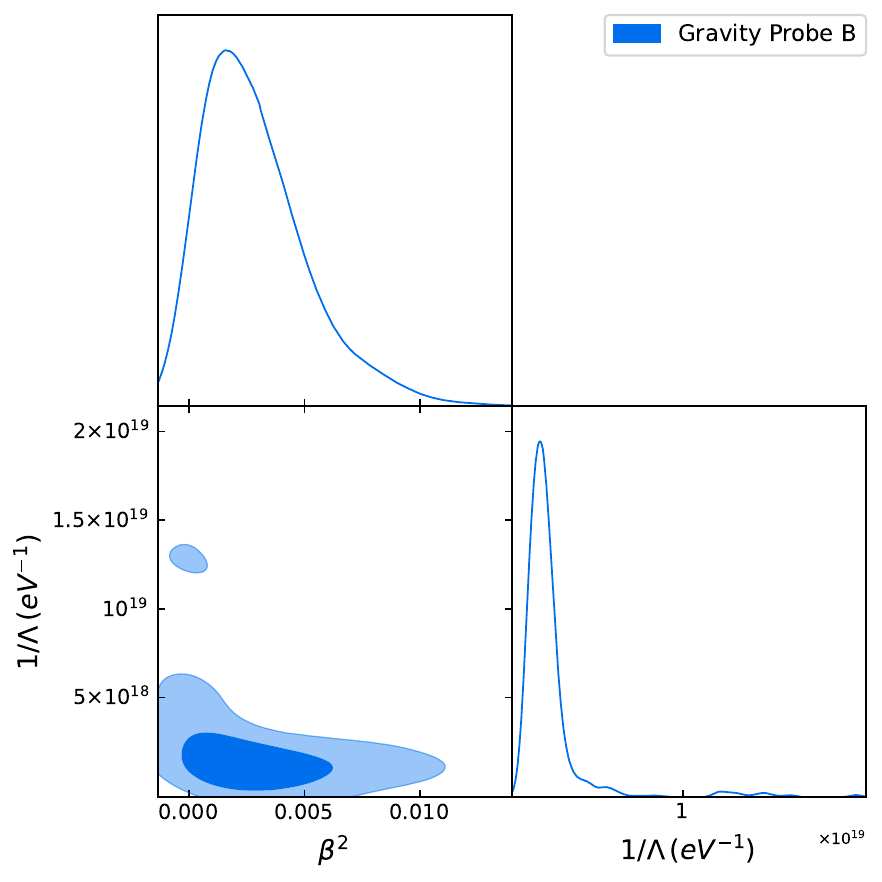}
\caption{{Bounds on the conformal and the disformal couplings from the Gravity Probe B experiment which tests the geodetic and the FD effects and therefore gives a bound on the conformal coupling $\left(2.963 \pm  2.045 \right) \cdot 10^{-3}$ and the disformal coupling $\Lambda > 3.92 \cdot 10^{-19} \rm{eV}$. }}
     	\label{fig:sat}
\end{figure}
\begin{figure}
 	\centering
  \includegraphics[width=0.46\textwidth]{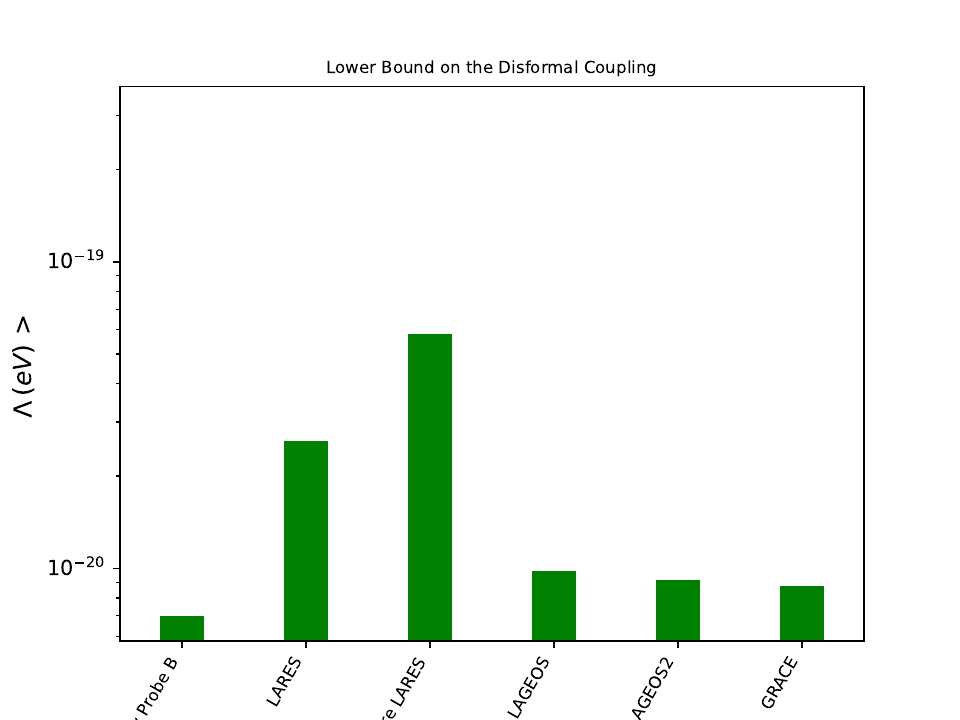}
     \caption{\it{The limit on the disformal coupling from the satellite experiments complemented with the Cassini bound on the conformal coupling. The range is compatible with $\Lambda > 10^{-19}\, \rm{eV}$.   }}
     	\label{fig:SatLam}
\end{figure}

\subsection{Gravity Probe B}
Gravity Probe B (GPB) was a satellite-based experiment designed to test the geodetic and the FD effects. This was to be accomplished by measuring very precisely tiny changes in the direction of the spin of four gyroscopes contained in an Earth-orbiting satellite at $650 \rm{km}$. Ref.~\cite{Everitt:2011hp} reports that analyses of the data from all four gyroscopes result in a geodetic drift rate of $-6601.8\pm 18.3 \, \rm{mas/yr}$\footnote{$\rm{mas}$ stands for milli-arc-second.} and a FD drift rate of $-37.2 \pm 7.2 \rm{mas/yr}$, in good agreement with the general relativity predictions of $-6606.1 \pm 0.28 \, \rm{mas/yr}$ and $-39.2 \pm 0.19 \, \rm{mas/yr}$, respectively.

This provided a way to test different theories of gravity such as Yukawa type potentials ~\cite{Capozziello:2014mea}, Horava-Lifshitz gravity~\cite{Radicella:2014jwa}, light scalars or pseudoscalars coupled to leptons and affect the precessional \cite{Poddar:2021ose}. and the first constraints on conformal and disformal interactions ~\cite{Brax:2020vgg}. Since the orbital radius of the satellite is $7027.4 \, {\rm km}$, the orbital period is $0.0107 {\rm Hz} = 7\cdot 10^{-19} \, \rm{eV}$. The posterior distribution of GPB is presented in Fig~\ref{fig:sat}. The complete MCMC yields a fit of
\begin{equation}
\beta^2 = \left(2.963 \pm  2.045 \right) \cdot 10^{-3}
\end{equation}
on the conformal coupling, giving $\Lambda > 3.92 \cdot 10^{-19}\, \rm{eV}$ on the disformal coupling. The result is compatible with GR at the  $2\sigma$ level. Taking the Cassini bound on the $\beta$ parameter gives a bound of $\Lambda >  5.33  \cdot 10^{-21} \, \rm{eV}$ on the disformal coupling. Notice that this is much higher  than the Hubble rate now $H_0\simeq 10^{-33}$ eV which would correspond to a suppression scale of the disformal coupling at the dark energy scale \footnote{The preferred value for disformal theories with an effect on the dynamics of the Universe is $\Lambda\simeq 10^{-33}$ eV. This follows from the presence of extra terms in the cosmological equations in $\partial_t/\Lambda$, i.e. time derivatives suppressed by the cut-off scale $\Lambda$. Typically one expects dynamical effects from the scalar field when these terms are of order unity. Moreover as the scalar field also evolves on time scales of the order of the age of the Universe $H_0^{-1}$ at late times, i.e. when Dark Energy plays a role, this is only possible for a cut-off scale $\Lambda \simeq H_0$. As can be seen, this regime with  a low cut-off scale is on the verge of the admissible energy range for a low energy effective field theory as higher order terms in the derivative expansion of the disformal term might be required.  }

\subsection{LARES, LAGEOS and GRACE}
The Laser Relativity Satellite \texttt{(LARES)}\footnote{\url{ https://www.asi.it/scienze-della-terra/lares/}}  was launched to measure the FD effect with an accuracy of about $10^{-2}$ \cite{1998Sci...279.2100C,2009SSRv..148...71C,Lucchesi:2019kwm,Capozziello:2014mea}. The body of this satellite has a diameter of about 36.4 cm and weighs about 400 kg. The satellite was set on an orbit with an altitude of $1450 {\rm km}$, an inclination of $69.5 \pm 1$ degrees and eccentricity $9.54 \cdot 10^{-4}$. Tests of the FD precession consist of small secular precessions of the orbit of a test particle in motion around a central rotating mass. For example, this has  been performed with the LAGEOS satellites~\cite{Ciufolini:2004rq} where the satellite acts as the particle moving around the earth. The orbital period of these systems is about $9\cdot 10^{-4} {\rm Hz} = 6\cdot 10^{-19} \, \rm{eV}$. Since these experiments constrain the FD effect we complement them with the Cassini bound on  the conformal coupling, and from these satellite experiments we get a range of $\Lambda >  10^{-20} \, \rm{eV}$ on the disformal coupling. Fig~\ref{fig:SatLam} summarises the different satellite experiments with the different constraints. Notice that all these experiments do not exclude the dark energy scale as a suppression scale for the disformal interaction. 

\subsection{Gravity Probe Spin}
In~\cite{Fadeev:2020gjk} was suggested that future measurements of relativistic FD and geodetic precessions should use  the intrinsic spin of the electron, hence called Gravity Probe Spin (GPS). Such a measurement would be  possible by using ${\rm mm}$ scale ferromagnetic gyroscopes in orbit around the Earth. Fig~\ref{fig:GPS} shows the lower bound on the disformal coupling vs. the future measurement error of the GPS experiment which is order of $10^{-18} \,$ eV.

\section{GINGER}
\label{sub:GINGER}
GINGER (Gyroscopes in General Relativity) relies on the  difference in time of flight of two counter propagating waves in a closed path, i.e. the Sagnac effect~\cite{Ruggiero:2015gha,Tartaglia:2016jfo,DiVirgilio:2017fuh,DiVirgilio:2020ior,Bosi:2020dgg,DiVirgilio:2021ziy,Altucci:2023wdo}. The effect depends on the lack of reciprocity of the two directions and is related to the FD effect introduced by a rotating object. The difference in the time of flight is generated by the  Ring Laser Gyros which emit these counter propagating waves. GINGER will measure the difference of time of flight  with an accuracy down to $\sim 10^{-4}$ that will be used to test GR and other theories of gravity ~\cite{Capozziello:2021goa}. For the generic metric of a rotating gravitational object
\begin{equation}\label{eq:metric_rot}
    ds^2 = g_{00}dt^2+2g_{0i}dtdx^i+g_{ij}dx^idx^j
\end{equation}
null geodesics  are given by 
\begin{equation}
    dt= -\frac{g_{0i}}{g_{00}}dx^i -\frac{1}{g_{00}}\left((g_{0i}dx^{i})^2-g_{00}g_{ij}dx^idx^j\right)^{1/2}.
\end{equation}
Parameterising the path that light follows in space in terms of a parameter $l$, and assuming a closed path of  circumference $P$, i.e.  $x^i(l) = x^i(l+P)$, we have the equations of the trajectory
\begin{equation}
    \frac{dt}{dl}= -\frac{g_{0i}}{g_{00}}\frac{dx^i}{dl} -\frac{\epsilon}{g_{00}}\left((g_{0i}\frac{dx^{i}}{dl})^2-g_{00}g_{ij}\frac{dx^i}{dl}\frac{dx^j}{dl}\right)^{1/2}
\end{equation}
where  $g_{00}<0$ and $dl= \epsilon \vert dl \vert$ with our choice of signature. 
We are interested in sending photons along the closed path in the two opposite directions with  ($dl>0$) and ($dl<0$) respectively. The proper  time delay between these two trajectories is given by
\begin{equation}
    \Delta \tau = -{2}\sqrt{-g_{00}}\oint \frac{g_{0i}}{g_{00}}\frac{dx^i}{dl} dl
\end{equation}
The scalar field background influences this time delay as the metric considered here is the Jordan metric. Using the small field expansion, we get the leading order effect
\begin{equation}
\Delta \tau = {2}\oint h_{0i}^J\frac{dx^i}{dl} dl= -\frac{4}{m^2_{\rm Pl}\Lambda^2} \oint \partial_0 \varphi \partial_i \varphi\frac{dx^i}{dl} dl
\end{equation}
which only involves the disformal coupling. As an example, a closed loop at the surface of a body  considered as a    test body in the field of a larger one, e.g the Earth with its orbit around the Sun,  will give rise to 
a time difference. Let us expand the field
\begin{equation}
\varphi(\bar x+ x(l))= \bar \varphi + x^i(l) \partial_i \bar \varphi+\dots
\end{equation}
where $\bar x$ is the centre of the loop whilst $\bar\varphi$ and its derivatives are their values at the centre of the loop. The first non-vanishing contribution to the time delay is given by
\begin{equation}
\Delta \tau =  \frac{4}{m^2_{\rm Pl}\Lambda^2} ( \partial_{0}\partial_{j}\bar \varphi \partial_i \bar \varphi+\partial_{0}\bar \varphi \partial_i\partial_j \bar \varphi)\oint x^j\frac{dx^i}{dl} dl .
\end{equation}
Now for closed planar loops we have
\begin{equation}
\oint x^j\frac{dx^i}{dl} dl= A \epsilon^{jik}n_k
\end{equation}
where $n_k$ the unit vector orthogonal to the loop and $A$ its surface area. 
With this we obtain the contribution of the disformal interaction to the time delay to be
\begin{equation}
\frac{\Delta \tau}{A} =  \frac{4}{m^2_{\rm Pl}\Lambda^2} \epsilon^{ijk}  \partial_{0}\partial_{i}\bar \varphi \partial_j \bar \varphi
\end{equation}
where $A$ is the area encircled by the light beams. In the case of a loop at the surface of the Earth in the background of the Sun which is static in first approximation this becomes
\begin{equation}
\frac{\Delta \tau}{A} =  \frac{128\pi G^2 m_\odot m_\oplus \beta^2}{\Lambda^2 r^3 d^3}\epsilon^{ijk}\left(v_i - 3\frac{ ( v^jr_j)}{r^2}r_i\right) d^j n^k.
\end{equation}
where $\vec v$ is the velocity at the loop comprising the effects of the Earth's velocity in the heliocentric frame and the Earth's rotational velocity, $\vec r$ the position of the centre of the loop on Earth and $\vec d$ the position of the Earth compared to the Sun. Using vectorial notation we have:
\begin{equation}
\Delta \tau_{\text{dis}} =  \frac{128\pi G^2 m_\odot m_\oplus \beta^2}{\Lambda^2 r^3 d^2}\left((\vec v - 3v_r \vec e_r)\times\bar{d}\right)\cdot  \vec{A}
\end{equation}
where $v_r= \vec v . \vec e_r$ and $\vec e_r= \frac{\vec r}{r}$.
As expected, the time delay scales with the surface area of the loop and involves the projection of the angular momentum of the Earth around the Sun $\vec d\times \vec v$ along the normal to the loop $\vec n$. Finally we denote by $\vec A= A\vec n$ the surface vector pointing in the normal direction to the loop. Based on~\cite{Bosi:2011um} and performing the calculation in linear approximation for an instrument with its normal contained in the local meridian plane, the GR result is 
\begin{eqnarray}
&&
\frac{\Delta \tau_{GR}}{4 \Omega_{\oplus} A} =  \cos(\theta + \alpha) - 2 \frac{G M}{R_\oplus c^2} \sin \theta \sin \alpha 
\nonumber \\ 
&& + \frac{G I_{\oplus}}{R_{\oplus}^3 c^2}\left( 2 \cos \theta \cos \alpha + \sin \theta \sin \alpha\right).
\end{eqnarray}
where $\alpha$ is the angle between the local radial direction and the normal to the plane of the instrument, measured in the meridian plane, $\theta$ is the colatitude of the laboratory, and  $\Omega_{\oplus}$ is the rotation rate of the Earth as measured in the local reference frame. 

\begin{figure}
 	\centering
  \includegraphics[width=0.44\textwidth]{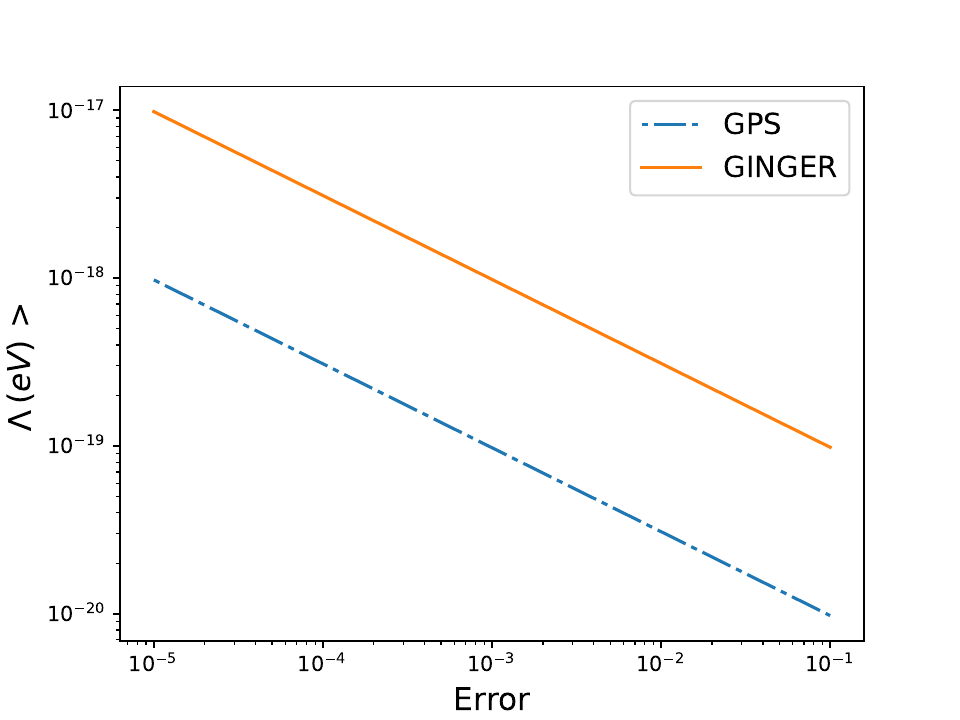}
     \caption{\it{The limit on the disformal coupling from the future GINGER and the Gravity Probe Spin  experiments in addition to the Cassini bound on the conformal coupling. The range is about $\Lambda > 10^{-17}\, \rm{eV}$ for GINGER and $\Lambda > 10^{-18}\, \rm{eV}$ for the Gravity Probe Spin.   }}
     	\label{fig:GPS}
\end{figure}

In order to determine the contribution of $\Delta \tau_{\text{dis}}$ we focus of the partial ratio, that gives
\begin{equation}
\frac{\Delta \tau_{\text{dis}}}{\Delta \tau_{\text{GR}}} \sim \frac{32\pi G^2 m_\odot m_\oplus \beta^2 v_{E}}{\Omega_{\oplus} \Lambda^2 R_{\oplus}^3 d^2}.
\end{equation}
Since the future error of the GINGER experiment should be around $10^{-4}$ and taking the Earth velocity of order $30~{\rm km/sec}$, the lower limit on $\Lambda$ should be around $ >\, 3.1 \cdot 10^{-17} \, \rm{eV}$. In this case the dark energy scale would be strongly disfavoured as a suppression scale for the disformal coupling. Fig~\ref{fig:GPS} shows the lower bound on the disformal coupling vs. the future measurement error of the GINGER experiment. This scaling assumption is compatible with the ladder expansion since the velocity we discuss here is much lower then the speed of light.

\begin{figure}
 	\centering
\includegraphics[width=0.47\textwidth]{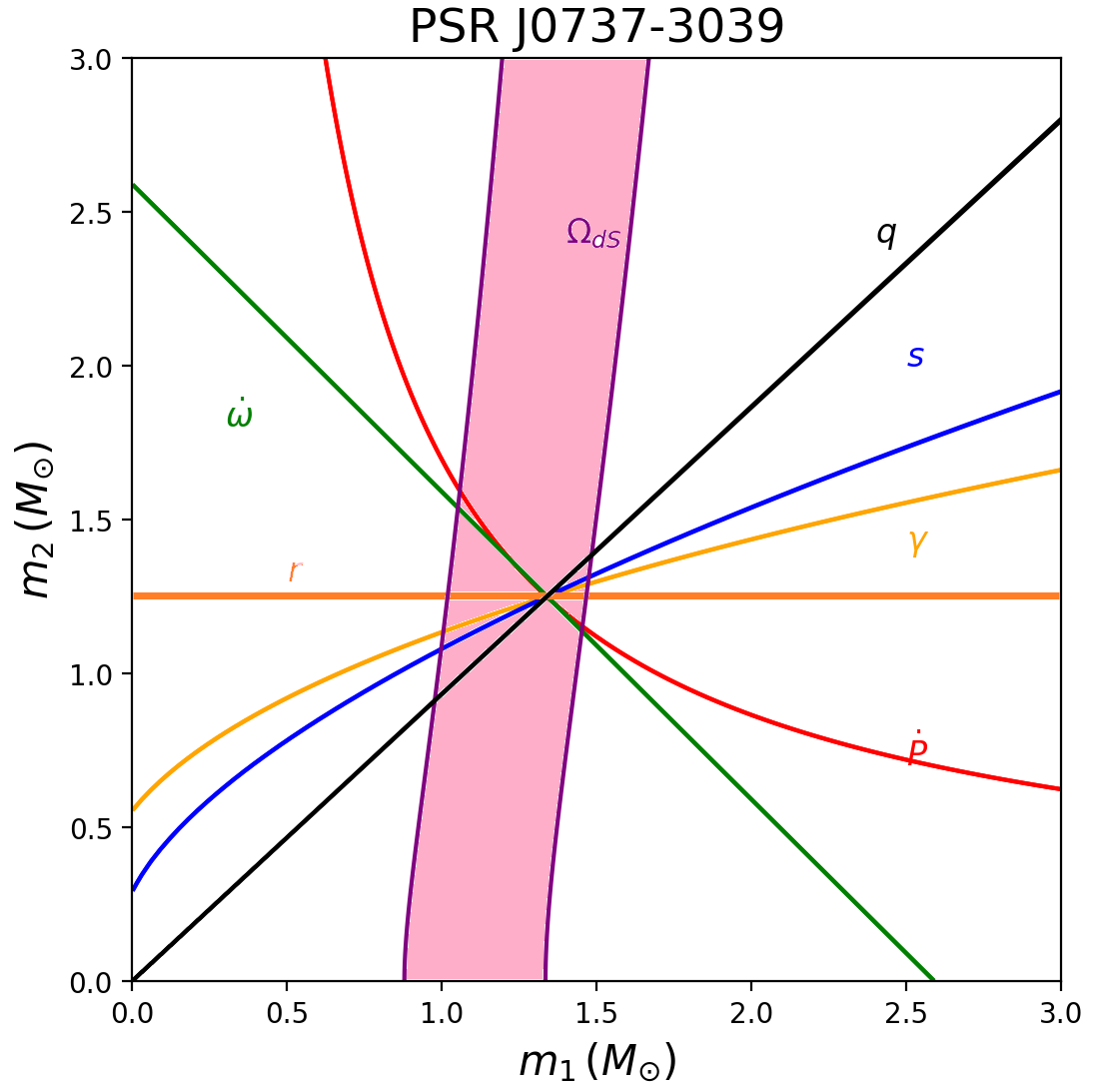} 
     \caption{\it{ The mass-mass diagram of the double pulsars PSR J0737-3039 A/B with the post Keplerian Parameters. The contour describes the Post Keplerian Parameters and the width of each curve indicates the measurement uncertainty of the corresponding parameter. }}
     	\label{fig:doublePulsar}
\end{figure}
\section{Pulsars}
\label{sec:pulsars}
So far we have only considered satellites in the Earth's atmosphere. We change environment and consider precession effects further in the Milky Way where signals from pulsars have been observed. A pulsar is a highly magnetised rotating neutron star that emits radiation from its magnetic poles. This radiation can be observed only when a beam of emission is pointing towards the Earth (similar to the way a lighthouse can be seen only when the light points  in the direction of an observer), and is responsible for the pulsed appearance of emissions. Binary pulsars are one of the best systems in astronomy in order to measure the Post Keplerian Parameters (PKP) such as the orbital period decay $\dot{P}$ and the periastron advance $\dot{\omega}$. Ref.~\cite{Benisty:2022lox} constrains light scalars  with conformal and disformal interactions from the PKP. In the following we will assume that the conformal coupling in the pulsar's environment is the same as in the solar system. This will allow us to impose the Cassini bound on the conformal coupling when computing the posterior distribution of the conformal and disformal couplings. 

The PKP which are accessible from the pulsar timings are the  Einstein $\gamma_E$ parameter accounting for the time delay due to the time dilation and the gravitational redshift of the pulsar signal in the solar system, the Shapiro time delays due to the crossing by the signal of the potential well of the solar system (this includes both the Shapiro delay shape $s$ and the Shapiro delay range $r$, see \cite{Benisty:2022lox} for their definition):
\begin{equation*}
\gamma_E =  e \, m_c \, \sqrt[3]{\frac{T_{\odot}^2}{ n_b m}  } \left( 1+ 2 \beta^2\right)^{2/3} \left(1 + \frac{m_c}{m}\right) ,    
\end{equation*}
\begin{equation}
s = \frac{x_p}{m_c} \sqrt[3]{\frac{n_b}{ 1+ 2 \beta^2} \frac{m^2}{T_{\odot}}},  \quad r = \left(1+ 2 \beta^2 \right) T_{\odot} m_c,  
\end{equation}
with the gravitational wave emission rate:
\begin{equation}
\dot{P} =  -\frac{195\pi T^{5/3}_{\odot}}{5 n_b^{5/3}}\frac{m_p m_c}{m^{1/3}}\left[(1+\frac{\beta^2}{3})f_1(e)+\frac{10}{9} \beta^2 \, f_2(e) - \epsilon_{\Lambda} \frac{20}{3} f_3(e), \right],  
\label{eq:pbdot}
\end{equation}
where:
\begin{equation*}
f_1 (e) = \frac{1+\frac{73}{24}e^2+\frac{37}{96}e^4}{(1-e^2)^{7/2}}, \quad 
f_2(e) = e^2 \frac{1+\frac{1}{4} e^2}{(1-e^2)^{7/2}}  
\end{equation*}
\begin{equation*}
f_3(e) = e^2\frac{1+\frac{37}{4}e^2+\frac{59}{8}e^4 + \frac{27}{64}e^6}{(1-e^2)^{13/2}}. 
\end{equation*}
Finally we also include the periastron advance
\begin{equation}
\begin{split}
 \dot{\omega} = \left(m T_\odot \right)^{2/3}\frac{n_b^{5/3}}{1 - e^2} \left[3 - 2\beta^2  + \frac{5\epsilon_\Lambda }{2 \pi T_{\odot} \Lambda^2} \right].
\end{split}
\label{eq:prec}
\end{equation}
Here $m_p$ is the pulsar mass, $m_c$ is the companion mass, $m = m_p + m_c$ is the total mass of the system and  $x_p$ is the projected semi-major axis.  These PKP provide a significant test of the conformal and disformal interactions  leading to stringent constraints on the couplings of light scalars to matter.   

\begin{figure}
    \centering
    \pgfmathsetlength{\imagewidth}{\linewidth}%
    \pgfmathsetlength{\imagescale}{\imagewidth/524}%
    \begin{tikzpicture}[x=\imagescale,y=-\imagescale]
        \node[anchor=north west] at (0,0) {\includegraphics[width=\imagewidth]{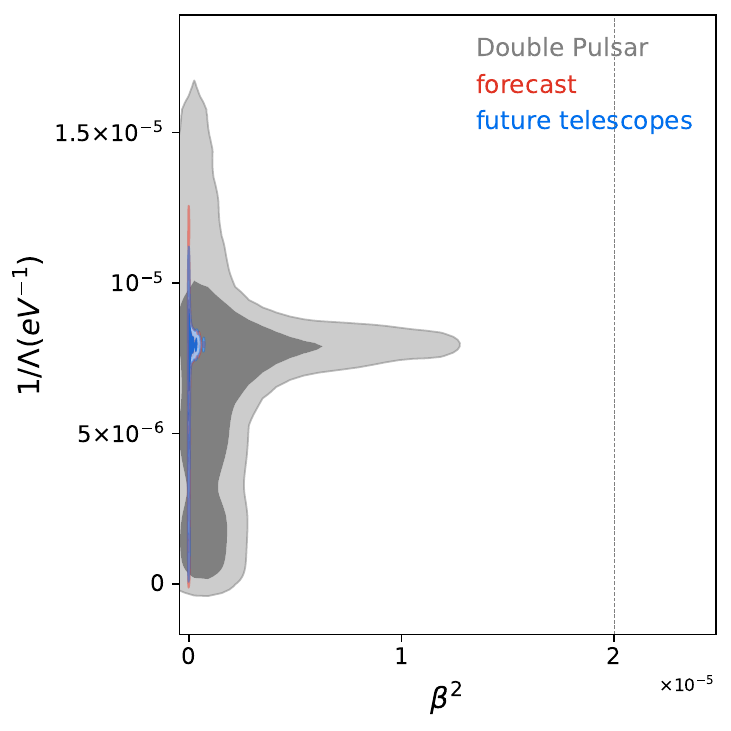}};
        \node[anchor=north west] at (250,160) {\includegraphics[width=0.5\imagewidth]{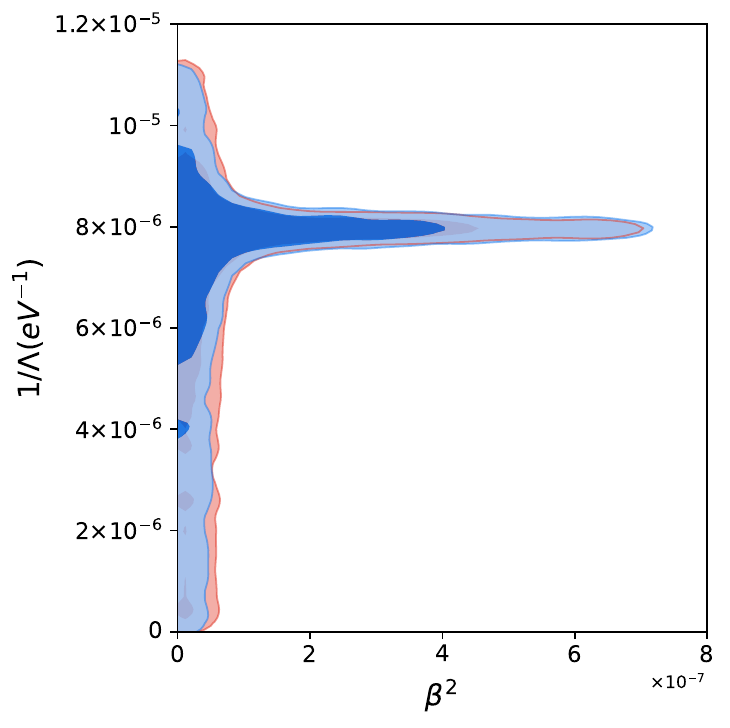}};
    \end{tikzpicture}
\begin{tabular}{| c | c | c |}
\hline\hline
 & $\beta^2$ & $\Lambda \, (MeV) > $ \\
\hline\hline
Current & $\left(1.94\pm 0.72\right) \cdot 10^{-5}$ & $1.62$\\
\hline
Forecast & $\left(1.16 \pm 1.84\right) \cdot 10^{-7}$ &2.1\\
\hline
future telescopes & $\left(0.99 \pm 1.53\right) \cdot 10^{-7}$ & 3.0\\
\hline\hline
\end{tabular}
\caption{\bf \it{The posterior probability distribution for the conformal and the disformal couplings (with $1 \sigma$ and $2\sigma$ contours) after taking into account the measurements from PSR J0737-3039 A/B (grey). The forecast for future constraints on the conformal and the disformal interactions are given from PSR J0737-3039 A/B (red) and  including  future telescopes (blue). For future measurements, the covered area reduces and the upper bound on the conformal coupling and the lower bound on the disformal coupling change. }}
\label{fig:doublePulsarPost}
\end{figure}
In this paper we are interested in precession effects. It turns out that relativistic geodetic effects  were detected and constrained using different binary pulsars. For instance the pulsar PSR J1141-6545 gives results for the geodetic effect \cite{Manchester:2010dh,VenkatramanKrishnan:2019xpt}. PSR J0737-3039 is a double pulsar ~\cite{Kramer:2021jcw} that gives a direct value for the geodetic precession \cite{Breton:2008xy,Kramer:2010hd}. We include the geodetic precession in the likelihood analysis that we perform in order to constrain the masses $m_p$ and $m_c$ together with the couplings $\beta$ and $\Lambda$. The PKP involve the  four unknown quantities $m_p, m_c, \beta, \Lambda$ which should be extracted from the observables $n_b, e, x_p, r, s, \dot{P}_b$. This can be achieved from the likelihood:
\begin{equation}
-2\ln \, \mathcal{L} \left(m_p,m_c,\beta,\Lambda \right) = \sum_{i = 1}^{N_{PSR}} \left(\frac{\xi(m_p,m_c,\beta,\Lambda) - \xi_{ob}}{\delta \xi_{ob}}\right)^2
\end{equation}
where $\xi$ is one of the corresponding PKP taken from the list $\xi \in [\dot{\omega}, \dot{P}, \gamma_E, r, s, q, \Omega_{dS}]$ with the error $\delta\xi$. Here $q$ is the ratio of the masses $q = m_p/m_c$. The prior we consider for the PKP are Gaussian priors as reported in the original papers. For the masses we put a uniform prior of $[0,3] M_{\odot}$. Since the conformal interaction could be present without the disformal interaction, we test two different cases: only the conformal interaction and the the conformal with the disformal interaction.

Fig~\ref{fig:doublePulsarPost} shows the posterior probability distribution for the conformal and the disformal interactions from  two different analyses. As the PKP depend on the masses of the pulsars and the companion star, the conformal and the disformal interactions, one has to use at least four PKP to extract constraints from data. The table and Fig~\ref{fig:doublePulsarPost} shows the resulting constraints  for the scalar interactions. We include in our analysis the de-Sitter precession  $\Omega_{dS}$. One can see that the resulting bounds are  strong and comparable to the Cassini bound (the grey line): $\beta^2 = \left(1.939\pm 0.724\right) \cdot 10^{-5}$ and $\Lambda > 1.62\,$ MeV. This result is compatible with GR at the $2\sigma$ level.

Fig~\ref{fig:doublePulsar} shows the mass-mass diagram of the double pulsar. Any two lines give the contour of the corresponding PKP (with a $1 \, \sigma$ error) for different masses (the pulsar mass vs. the companion star). In this case of coupled scalars, we include the best values of the conformal and the disformal interactions. Since the contours intersect at  the same point in the mass-mass diagram, the model predicts the masses of the two pulsars  and bounds the  conformal and the disformal interactions up to the limit of the posterior values.

Ref.~\cite{Kehl:2016mgp} states that with additional years of timing measurements and new telescopes like the Square Kilometre Array (SKA) and others, the precision of these tests will increase and new effects like the FD precession of the orbit will become measurable. In this way, one could distinguish between the precession $\dot{\omega}$ and the FD precession $\dot{\omega}_{FD}$ giving stronger constraints on the conformal and the disformal interactions.

Fig~\ref{fig:doublePulsarPost} shows the future constraint on the interactions using the forecast from Ref.~\cite{Kehl:2016mgp}. Ref.~\cite{Kehl:2016mgp} uses simulations for future constraints with or without other telescopes to  reduce current uncertainties. We use the future error that Ref.~\cite{Kehl:2016mgp} estimates to be within reach in $2030$ for different PKP. Future constraints should improve the bounds on the conformal and the disformal interactions, i.e.  the conformal interaction upper bound will be at the $10^{-6}$ level and $10^{-7}$ when other  telescopes are taken into account. The bound on the disformal interaction will be of  the same order ($\sim {\rm MeV}$) but stronger when  other telescopes are taken into account.

\section{S stars in the Galactic Centre}
\label{sec:center}

\begin{figure}
 	\centering
\includegraphics[width=0.47\textwidth]{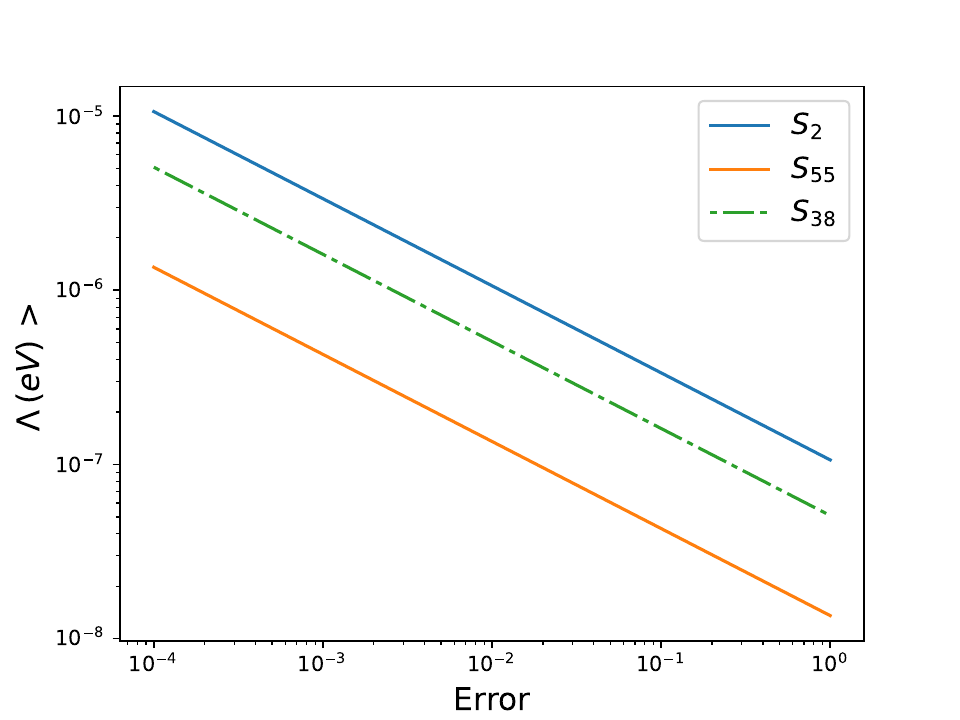} 
     \caption{\it{Forecast of the lower bound on the disformal coupling from the frame dragging effect on some S-stars in the galactic centre combined with the Cassini bound on $\beta$. }}
     	\label{fig:Sstarsupperbound}
\end{figure}

The centre of the Milky Way hosts the closest supermassive black hole, Sgr A*. The stars orbiting $Sgr A^{*}$ are called  S-stars ~\cite{Grould:2017bsw,Yu:2016nzn,2016ApJ83017B,Abuter:2018drb,2009ApJ707L114G,Do:2019txf,Abuter:2020dou,Amorim:2019hwp,2017ApJ...845...22P,Benisty:2021cmq,Ben-Salem:2022txj} with decades of monitoring  of their locations and velocities. A large fraction of these stars have orbits with high eccentricities. Thus, they reach high velocities at the pericentre and can be used for constraining scalar interactions. There are a few studies that discuss the FD precession in the S-stars motion \cite{Kannan:2008nm,2011Msngr.143...16E,Zhang:2015nya,Iorio:2017fck,Iorio:2017med,Fragione:2022oau}. Ref.~\cite{Iorio:2020uet} claims that the FD effect is overwhelmed by the systematic uncertainties in the Schwarzschild parameters due to the current errors in the stars’ orbital parameters and the mass of Sgr A* itself. \cite{Merritt:2009ex,Grould:2017bsw} claims that detection of FD precession may be feasible after a few years’ monitoring with an instrument like GRAVITY for orbits of some S-stars. Especially, the S2 star orbits with a period of 16 years and it should be possible to constrain the angular momentum of the black hole by observing the star over 32 to 48 years. Ref.~\cite{Fragione:2022oau} claims inconsistency between the current measurements of the Event Horizon Telescope predictions for the Sgr-A* spin and the bound from the S-stars orbits.

Using the known properties of the $S$ stars (their masses and periods as in Ref.~\cite{2017ApJ...837...30G}) we forecast a bound on the disformal coupling , which depends on the errors of the future measurements of the quantities appearing in Eq~(\ref{eq:modFD}), ie the mass of the S-stars, the eccentricity and the angle $\psi$. Fig.~\ref{fig:Sstarsupperbound} shows that we obtain a  lower bound around $\Lambda >  10^{-5} \, \rm{eV}$ after imposing the Cassini bound  on the conformal coupling. We use the known masses of these S-stars and the predicted accuracies of the future measurements. Other S-stars around the galactic centre have shorter periods than the S2 star, such as the  S4711, S62, S4714 or S4716~\cite{Iorio:2020uet,2022ApJ...933...49P} and will give stronger constraints on the disformal interaction.  However, the masses of these stars is still unknown. This forecast only applies if the supermassive black hole at the centre of the galaxy has a scalar charge. This could be the result of a violation of the no-hair theorem by the time dependence intrinsic to both the galactic and cosmological environments, see \cite{Wong:2019yoc} for instance.

\section{Discussion and Summary}
\label{sec:results}
In this paper we investigated the consequences  of a light scalar coupling to matter  on the geodetic and the FD effects.  Both conformal and disformal couplings of the scalar field to matter are considered and used to generate geodetic and FD effects. To first order in post Newtonian expansion, the correction to the solution of the scalar field was obtained in ~\cite{Brax:2018bow}. This was extended in ~\cite{Benisty:2022lox} to  the Post Keplerian Parameters with conformal and disformal interactions, enabling our current study.

Eq.~(\ref{eq:modDS}) and~(\ref{eq:modFD}) shows the relative modification to the geodetic and the FD precessions respectively. If only the conformal interaction is present, then the  geodetic effect is modified whilst the FD effect is affected  only if both the conformal and the disformal couplings exist. The geodetic effect gives constraints on $\beta$ directly and the FD effect gives constraint on $\beta/\Lambda$. For the  experiments that measure the FD effect directly, we use the Cassini spacecraft bound on the conformal coupling as a prior, and deduce  constraints on the disformal coupling.  

The bounds on the coupling obtained from satellite experiments are strictly solar system constraints. This is why we can complement them with the Cassini bound. We find a bound on the disformal scale of order $\Lambda \gtrsim 10^{-18}$ eV which is much smaller than the one from pulsar timing $\Lambda \gtrsim 1$ Mev or even particle physics $\Lambda\gtrsim 650$ GeV. As the energy scales and the environments involved in pulsars and particle colliders are very different from the earth's atmosphere, we simply notice that a strong variability with the environment is allowed for the disformal coupling. 

The satellite-based experiments measure directly the FD effect while in the case of pulsars the effect is derived from the pulses sent to earth. Since the disformal interaction depends on the period, different systems with different periods will give different bounds. However, the satellite experiments have the advantage of measuring the FD effect directly.

The strongest bound on the conformal and disformal couplings from the geodetic effect is from the precession of binary pulsars and especially from the double pulsar~\cite{Benisty:2022lox}. The current observations of the double pulsar gives a known bound on the geodetic precession value and in the near future one will be able to measure the FD precession directly~\cite{Kramer:2010hd}. Future measurements will help distinguishing between the first and the second Post Newtonian contributions to the precessions and  the FD contribution. With this separation the constraints on the disformal coupling should be more stringent with the increased  precision on both  $\dot{\omega}$ and $\dot{\omega}_{FD}$.  This will allow for a stronger test of light scalar couplings combining pulsar timing and precession effects.

\acknowledgements 
We would like to thank to Scott Melville, Leong-Khim Wong and the anonymous referee for useful comments and discussions. ACD thanks Chandrima Ganguly for discussions and collaboration at an early stage of this work. D.B. thanks Pasha Fadeev and Jenny Wagner for useful discussions. D.B. gratefully acknowledges the support of the Blavatnik and the Rothschild fellowships. D.B. acknowledges a Postdoctoral Research Associateship at the Queens' College, University of Cambridge. D.B. have received partial support from European COST actions CA15117 and CA18108 and the research grants KP-06-N58/5.

\bibliographystyle{apsrev4-1}
\bibliography{ref.bib}

\end{document}